\documentclass[10pt,twocolumn]{article}
\usepackage{times}

\PassOptionsToPackage{hyphens}{url}\usepackage[hidelinks]{hyperref}

\usepackage{xspace}
\usepackage{enumitem}
\usepackage{multirow}
\usepackage{graphicx}
\usepackage{xcolor}

\newcommand{\system}{FaaSFS\xspace}

\newcommand{\jss}[1]{}
\newcommand{\jmh}[1]{}

\begin{document}

\title{\Large \bf A FaaS File System for Serverless Computing}
\author{
Johann Schleier-Smith$^{1,2}$ \quad Leonhard Holz$^2$ \quad Nathan Pemberton$^1$ \quad Joseph M. Hellerstein$^1$ \\ [3mm]
\small {\em  $^1$UC Berkeley \quad
          $^2$Packet Computing}
}
\date{}
\maketitle


\begin{abstract}
Serverless computing with cloud functions is quickly gaining adoption, but constrains programmers with its limited support for state management.
We introduce a shared file system for cloud functions.
It offers familiar POSIX semantics while taking advantage of distinctive aspects of cloud functions to achieve scalability and performance beyond what traditional shared file systems can offer.
We take advantage of the function-grained fault tolerance model of cloud functions to proceed optimistically using local state, safe in the knowledge that we can restart if cache reads or lock activity cannot be reconciled upon commit.
The boundaries of cloud functions provide implicit commit and rollback points, giving us the flexibility to use transaction processing techniques without changing the programming model or API.
This allows a variety of stateful sever-based applications to benefit from the simplicity and scalability of serverless computing, often with little or no modification.
\end{abstract}

\section{Introduction}
\label{sec:introduction}

In some ways programming the cloud has never been easier---serverless computing puts the power of thousands of computers at developers' fingertips~\cite{jonas2017occupy,fouladi2019laptop}.
Autoscaling and pay-per-use mean that developers using Function as a Service (FaaS) platforms experience an illusion of infinite scale, and need not worry about allocating or administering the underlying resources~\cite{castro2019rise,jonas2019cloud}.
Still, serverless computing remains a new field, and users quickly run into limitations and difficulties~\cite{hellerstein2019}, especially when it comes to managing application \textit{state}.

Perhaps the most familiar and time-tested solution to state management is the POSIX file system API.
This led us to wonder about offering a POSIX filesystem API to serverless functions.
Our hypothesis was that the design patterns of cloud functions could offer opportunities for file system performance that could match or even exceed existing POSIX implementations for stateful VMs in the cloud.
In this paper we explore this design space.
Our Function as a Service File System (\system) implementation allows programmers to use familiar system interfaces, as well as existing software and libraries, while obtaining the scalability benefits of serverless computing and performance that is competitive with ``serverful'' NFS filesystems offered for cloud VMs.


\begin{figure}[!t]
  \centering
  \includegraphics[]{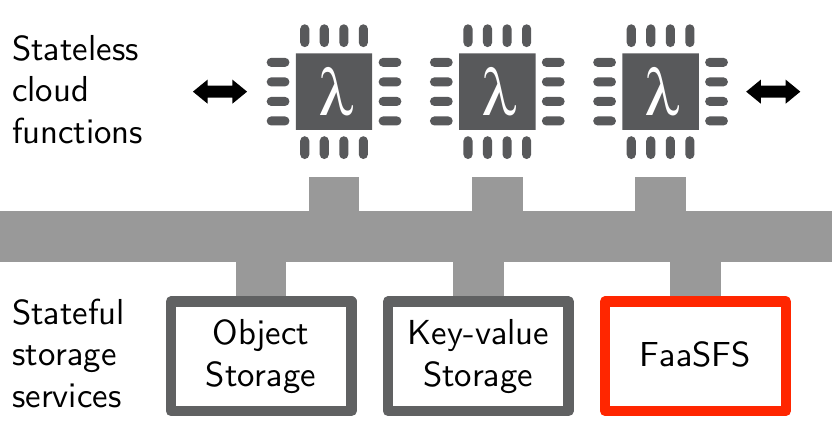}
  \caption{
  Serverless applications often use stateless cloud functions together with stateful services such as object storage or key-value storage.
  \system provides a shared file system for cloud functions. It provides a familiar POSIX API, and through caching and transactional isolation achieves performance near that of a local file system.
}
  \label{fig:tsfs-scaling}
\end{figure}

When using FaaS, also known as cloud functions, for serverless computing, programmers upload their applications as fragments of code written in high-level programming languages.
They then specify when these bits of code should run, e.g., in response to web requests or to events published on queues.
Cloud functions are \textit{stateless} in the sense that once a function finishes executing, the cloud provider can purge its execution environment, including the content of its local disk, and reassign resources to another customer.

Figure~\ref{fig:tsfs-scaling} illustrates a common solution to state management in serverless computing, which is to use a stateless FaaS tier with one of a range of stateful storage services.
Each cloud provider has different offerings,\footnote{Examples include S3 and DynamoDB on AWS, Storage and CosmosDB on Azure, and Cloud Object Storage and Datastore on Google Cloud.} which forces programmers to conform to provider-specific APIs.
This limits portability, and makes it particularly difficult to use serverless computing with existing code.

A shared POSIX file system API for the serverless cloud setting is easy to ask for, but can it be made to work?
Cloud functions are subject to all the vagaries of cloud execution~\cite{dean2013tail}, including resource contention and failures.
These challenges are exacerbated by a high degree of elasticity, which can raise parallelism from zero to thousands of concurrent executions in seconds~\cite{jonas2017occupy}.
One naturally concludes that the serverless environment is not conducive to implementing POSIX, which promises linearizability and has a chatty API that is vulnerable to latency.

\system processes POSIX calls \textit{optimistically}, using locally cached state and wrapping cloud function file system interactions in a transaction mechanism to recover consistency when conflicts ensue.
This approach is practical because programmers are accustomed to cutting up their applications into independent pieces when building cloud functions.
This style of programming is driven by two distinctive properties of the cloud functions environment:
\begin{itemize}[noitemsep,topsep=0pt]
\item \textit{Limited execution time}.
Cloud functions typically do one thing and then return, a property enforced by a configurable execution time limit (ranging from seconds to minutes).
\system can thus form transactions \textit{transparently} by starting and ending them at function boundaries; the programming model stays the same.
\item \textit{Function-grained fault tolerance}.
Cloud functions are routinely re-executed by the FaaS platform, and their programmer contract generally requires developers to write idempotent code. 
\system also requires programmers to write retry-safe code, but provides them with a transaction mechanism that makes it simpler to do so correctly.
\end{itemize}

Working with existing FaaS platforms is challenging because the environment is restricted in many ways, but we were are able to work around these problems.
Mapping POSIX into a transactional database context requires some care because the memory consistency guarantees that describe file systems do not correspond directly to the transaction isolation levels provided by databases.
Our implementation also needed to surmount several algorithmic challenges in order to achieve good performance: allowing high concurrency when the file length changes requires special treatment, and we utilize a fine-grained cache update mechanism (both described in Section~\ref{subsec:consistencyimpl}).

Our aim in this work is to demonstrate proof of design feasibility.
We focus on showing the performance necessary to run existing software, on integrating with the cloud functions environment, and on developing the protocols and mechanisms for ensuring transactional consistency and cache consistency.
None of these evaluations are exhaustive, and we leave construction of a fault tolerant and scalable back end to future work.
We believe, however, that the work presented here makes a good case for our approach and represents a full implementation of the salient serverless-specific aspects of our design.

Our evaluation demonstrates the value of POSIX compliance by running both real-world applications and synthetic benchmarks including Filebench and TPC-C.
By running a popular blog software, we also show how we can support an unmodified real world application.
Setting up this software with \system is much like running it in a local development environment, and involves none of the configuration necessary for a scalable and fault-tolerant server-based deployment.
For workloads that benefit from optimistic execution, \system can outperform traditional shared file systems accessed from statically allocated (serverful) VMs.
For all workloads, \system provides a bridge between a world of existing software and programming practice and the autoscaling and zero-configuration world of serverless cloud computing.

\section{Motivation}
\label{sec:motivation}


\subsection{Use cases}
\label{subsec:usecases}

Scalable image processing is the canonical example use case for serverless~\cite{baldini2017serverless}, and it illustrates the awkwardness of working with state.
While there are countless image processing tools that operate directly on files, these do not speak the various provider-specific APIs of cloud storage.
Using them in a serverless function involves rewriting them to first stage objects in local temporary storage, to then run the tools, and to then write outputs back to cloud storage, a process that introduces complexity.

Cloud functions also seem like a natural fit for web and API serving, but using them is not as simple as it might be because of state management.
FaaS offers an autoscaling model well matched to these use cases, and makes it simple to run web application code, which is typically stateless. However, even a simple web site such as a blog running on a single server will often use both a database and a local file system.
Meeting these state needs with cloud services is possible (see Figure~\ref{fig:tsfs-scaling}), however it requires programmers to use cloud-provider-specific APIs and may add complex configuration.
We return to this example in our evaluation of Section~\ref{subsec:fullstack}.

PyWren~\cite{jonas2017occupy} highlighted the benefit of cloud functions in making cloud computing accessible to a broad range of scientific computing users.
However, PyWren requires a customized distribution of Python designed for pre-loading into cloud functions.
This distribution is crafted to be sufficiently inclusive to be practically useful, while remaining small enough to fit into the limited storage space available on the cloud function.
The needs of this use case can be met more naturally with a shared file system.

\subsection{Limitations of shared file systems}
\label{subsec:sharedfilesystems}

In the cloud there are many storage systems that are superficially similar to a file system, but which for good reason do not provide a POSIX API.
POSIX makes a linearizability guarantee, which is attainable within a single server, but is costly in a distributed and fault-tolerant setting.
To take an early example, the designers of the Google File System~\cite{ghemawat2003google,McKusick2010-fa} chose weaker consistency guarantees, in favor of achieving greater robustness to server failures, in part because their workloads simply did not require stronger guarantees.
Object stores might look like file systems for the cloud, but typically store only immutable items.
Some, such as AWS S3 also provide only eventual consistency for many metadata operations.
Key-value stores allow programmers to select from a range of consistency guarantees (e.g., as in Dynamo~\cite{decandia2007dynamo} and AWS DynamoDB~\cite{dynamodb}).
They map names to bytes, as file systems do, however their APIs are more restrictive than POSIX, requiring key-level replacement for modification, rather than supporting updates or appends.

There are numerous shared file systems and protocols that promise POSIX compliance, or something close to it.
These include NFS~\cite{sandberg1985design,nfs4rfc}, SMB~\cite{smbprotocol}, Lustre~\cite{schwan2003lustre}, and GPFS~\cite{Schmuck2002-zp}.
These can work quite well in settings such a high-performance computing, where file transfer units are large and concurrent access is coarse-grained and controlled by a job framework.
They are more challenged in environments where concurrent access is common, e.g., as documented for NFS~\cite{kirch2006nfs}.
What all of these shared file systems have in common is that a client must either hold exclusive access to shared state (which could be a file, a directory, or a file range), or must wait on the network to perform operations on shared state maintained at the server.
The established approach to this is leases~\cite{gray1989leases}, but it is vulnerable to client failure or slowness\jmh{cite something to underwrite this vulnerability?}, and is inconsistent with the common assumptions for cloud infrastructure~\cite{dean2013tail}, which assume that failures and delays can be common.

\section{Requirements for \system}
\label{sec:design}

In designing \system we needed to simultaneously meet requirements for performance, POSIX conformance, and cloud function integration. We touch upon each in turn.

\subsection{Performance requirements}
\label{subsec:performance}

A key motivation for providing a file system API is compatibility with existing software ecosystems.
In practice this means  matching not only the API semantics, but also the performance characteristics.
Ideally, \system should deliver a service that works like a local disk, but in a distributed cloud setting.

The POSIX file system API is a chatty one, which makes this goal challenging and elusive.
Many applications still perform a positioned read by first performing a \texttt{seek} and then a \texttt{read}, even though the newer \texttt{pread} has now been available for over 20~years.
Some applications \texttt{open} and \texttt{close} files repeatedly, something that programmers see as relatively cheap but which every time induces access control checks along the entire path to root.
In common implementations, the \texttt{cp} command issues a sequence of writes with block size of 64~KB or less.
Compare that to the minium part size for multi-part uploads to AWS S3 object storage, which is 5~MB.

Even on a single server, kernel caching of data and metadata is essential to achieving acceptable file system performance.
In a shared file system setting caching is equally important, but becomes more difficult to achieve since clients are distributed.

Given these challenges, a key innovation in \system is the use optimistic execution to achieve greater performance than traditional shared file systems attain.
We attribute the advantages of \system{} to four benefits of its mechanisms:
\begin{itemize}[noitemsep]
    \item Optimistic lock elision---lock requests always succeed locally and speculatively, rather than traversing the network as in other shared file systems.
    Our commit validation phase ensures strict serializable transaction isolation \cite{herlihy1990linearizability}, which ensures that lock semantics are respected (see Section~\ref{subsec:posix}).
    \item Optimistic use of cached state---we serve read requests from local cache, speculatively assuming that this state will still be valid at commit time.
    \item Snapshot reads---we support multiversioned state, which allows readers to make progress irrespective of write activity.
    Our implementation achieves this through a pull-based cache update mechanism and with a multiversioned back end.
    \item Fine-grained cache updates---we update or invalidate client caches at a block level, whereas other systems operate at a file level. Our cache update mechanism takes advantage of a transactional backend, which retains the information necessary for block-level change tracking in the transaction log.
\end{itemize}
See Section~\ref{subsec:consistencyimpl} for additional discussion of these mechanisms.
Among the advantages listed above, optimistic lock elision and optimistic use of cache state require speculative execution capabilities.
Snapshot reads could be provided in a blocking transactional implementation, however the optimistic mechanism helps because there is no need to enter a snapshot mode explicitly; we can optimistically assume the transaction will be read-only.
Fine-grained cache updates are easy to provide once the change tracking needed for snapshots is in place, and could be incorporated into traditional shared file systems.

\subsection{Reconciling POSIX with transactions}
\label{subsec:posix}

Perhaps counterintuitively, our decision to use transactions in \system is motivated by performance.
The perennial challenge for shared file systems is overcoming network latency when enforcing locks and obtaining up-to-date file system state.
The use of optimistic transactions to transparently remove locks in programs has been established in the hardware context~\cite{rajwar2002transactional}.
We draw upon this approach, in addition to well established optimistic techniques for providing serializable isolation in databases~\cite{bernstein1981concurrency}.
Transactions are often used to simplify crash recovery or concurrent programming, but in \system these benefits are ancillary to their motivating purpose, which is to provide a mechanism for optimistic execution.

Databases traditionally offer one or more transactional guarantees~\cite{adya2000generalized}, whereas the behavior of file systems is dictated by the POSIX specification~\cite{to2018open}.
We address how the two can be reconciled after first reviewing each.

POSIX~\cite{to2018open} uses informal language in describing its guarantees, saying ``Writes can be serialized with respect to other reads and writes'' and ``If a read() of file data can be proven (by any means) to occur after a write() of the data, it must reflect that write()...A similar requirement applies to multiple write operations to the same file position.''
In more formal language we can understand this as requiring both \textit{atomicity} and \textit{linearizability}.
Atomicity ensures that POSIX operations are indivisible units, each of which must be seen to have happened in its entirety or not have happened at all.
This means, for example, that a \texttt{read} must never observe just part of a \texttt{write}.
Similarly, an object subject to a \texttt{rename} always appears with either the old name or the new one; it may not disappear or appear under both names, even transiently.
Linearizability is the \textit{consistency} property requiring that all operations on the file system reflect a single global total order, and that this order corresponds to real-time, as observed at each client using wall clocks~\cite{herlihy1990linearizability}; each operation must slot into the total order at a point between its observed start time and its observed completion time.

Transactions form envelopes around multiple operations and provide guarantees relating to these groups rather than to operations individually.
Transactional guarantees are commonly described by one of several \textit{isolation levels}, which originally described locking schemes but now have mechanism-agnostic definitions~\cite{adya2000generalized,crooks2017seeing}.
Alternatives to locking in databases include optimistic concurrency control~\cite{kung1981optimistic}, multiversioning~\cite{bernstein1981concurrency} and deterministic databases~\cite{thomson2010case}.
At lower isolation levels such as \texttt{read committed} or \texttt{repeatable read} a transaction may experience some effects of concurrent execution, whereas with \textit{serializable} isolation, each transaction always sees the database as if it were the only transaction executing.

How does serializable isolation compare to linearizability?
Database isolation guarantees alone make no promises about real-time behavior, as measured by observers comparing clocks, or even just running transactions one-by-one.
For example, a database retains the serializable guarantee for a read-only transaction even when evaluating it against a snapshot of its state at an arbitrary point in its past.
Preventing such behavior calls for real-time correspondence, which for databases is known as \textit{external consistency}~\cite{corbett2013spanner,gifford1981information}. \jmh{So (a) this is not database lit, and (b) I have never seen this paper and it's not online, and perhaps as a result I don't view it as a well-defined term--it seems to have gained some popularity because the spanner paper used it (though they provide their own definition). Strict Serializability seems to come from the Linearizability paper, and is in terms of precedence ordering, not wall-clocks.}
Serializable isolation with external consistency is also known as \textit{strict serializability}.
\jmh{Serializable isolation that respects a precedence order of transaction requests arriving in the system is known as Strict Serializability~\cite{herlihy1990linearizability}.}
\jss{need to come back to these definitions - esp. precedence order vs wall clocks. My read of POSIX is wall clocks. Will address in revision.}

When a transaction consists of a single operation, strict serializability is equivalent to linearizability~\cite{herlihy1990linearizability}, however this equivalence is not true in general.
In this paper, we demonstrate that transactions with strict serializability offer good semantics and performance in many use cases. But there are some tradeoffs due to the strong semantics.
To understand the limitations of wrapping cloud functions in serializable transactions, consider the fact that such functions are not able to communicate with one another through the file system since their updates are isolated from each other.
This precludes functionality that might be desirable, e.g., enabling pipelined processing of a file while it is still being written.
We defer exploration of other isolation models in \system to future work.

In sum, running cloud functions in a transactional wrapper and with strict serializability does not correspond directly to running them with shared access to a linearizable POSIX file system.
However, it is equivalent to running them \jmh{exactly once,} \jss{requires you to build dedup} one at a time, in sequence.
Cloud function applications get the semantics they expect from POSIX, provided they do not rely upon interactions with others running concurrently.

\subsection{Cloud function environment}
\label{subsec:lambdaenv}

The design of \system is heavily influenced by the characteristics of the cloud functions environment, which differs in several important ways from a traditional server.
This includes some limitations which are quite fundamental, as well as others that we accept only because we are not in a position to change the provider's infrastructure.
Key characteristics of the environment include:
\begin{itemize}[noitemsep]
\item \textit{No root privileges}. Lambda offers a controlled environment, and operations like mounting an NFS share are prohibited. We are also unable to mount FUSE user space file systems~\cite{fuse} or load kernel modules.
Thus we implement \system as a user space file server.
\item \textit{Function instances freeze between invocations}.
Lambda instances only run when they are processing a request invoked through the API.
In the frozen state no processing occurs, even if data arrives on the network or if a timer is scheduled to raise a signal.
Function instances retain \system cache state while frozen, but there is no way to update that state until the instance begins processing a new request.
Also, we cannot safely maintain delegations~\cite{howard1988scale} or leases~\cite{gray1989leases} across successive invocations because we may be unable to revoke them on demand.
\item \textit{No inbound network connections}.
Cloud functions live behind a NAT layer that prohibits inbound network connections.
While some workarounds have been demonstrated~\cite{fouladi2019laptop,wagner2019}, direct communication between functions is not part of the programming model.
Functions using \system must communicate through a shared back end instead.
\item \textit{No names for function instances}. While Lambda may create many instances of a cloud function (as dictated by load), there is no way to route an invocation to a particular instance.
Each invocation could go to any function instance, which prevents us from partitioning our cache.
\item \textit{Limited execution time}. See Section~\ref{sec:introduction}.
Cloud functions typically run for fractions of a second.
While they can run for minutes, we always have a bound on their execution time.
We use limited execution time to our advantage, creating transactions transparently and thus relieving programmers of using a separate transactional API.
\item \textit{Function-grained fault tolerance}. See Section~\ref{sec:introduction}.
Cloud functions must be safe to retry, a requirement that is usually met by requiring idempotent code.
Prior work on serverless key-value storage points out that transactional atomicity relaxes this requirement~\cite{sreekanti2020fault}: atomic visibility of updates ensures that all or none of a function's effects are visible in storage.
Hence the transactions provided by \system simplify idempotence for filesystem interactions, though the programmer must still take care to ensure the safety of side-effects outside of \system.
\end{itemize}

Our approach aligns with the final two characteristics, both of which are tied tightly to the cloud functions programming model.
Bounded execution is a signature attribute that sets cloud functions apart from serverful services.
Function-grained fault tolerance with retries is the simplest way to provide reliability.

We also rely upon the ability to retain cache state in the cloud function instance memory between repeated function invocations.
Even though this may be viewed as a defect in the FaaS model~\cite{Jangda2019-yc}, a concession on functional purity and a security threat vector, we believe that this it is here to stay.
For performance reasons, providers typically reuse function instances for many requests, only shutting them down and reclaiming them after they have been idle for minutes, or even hours~\cite{wang2018peeking}.
Starting up a new cloud function instance requires transferring over code, starting up a language runtime, and perhaps executing application-specific initialization, all before the invoked function can start executing.
This takes time, perhaps less than a second but also possible much longer~\cite{oakes2018sock}.
Some applications may load external data sets, and JIT language runtimes like JVMs maintain substantial internal state to improve performance.
As a consequence, \system is not alone in benefiting from cached state, cloud providers and applications benefit as well.


\section{Implementation of \system}
\label{sec:implementation}

We chose to implement \system from scratch, rather than modifying an existing file system or database implementation, or building on top of one.
This choice is important to achieving our aims because it allows us to integrate the caching mechanism with the concurrency control mechanism, which we do by using the transaction log as a source of updates for intermittently connected cloud function clients.
We break this review of the implementation into two parts, a discussion of the functionality of each component, and a discussion of state management and transactions.

\subsection{\system Components}

\begin{figure}[!t]
  \centering
  \includegraphics[scale=1.0]{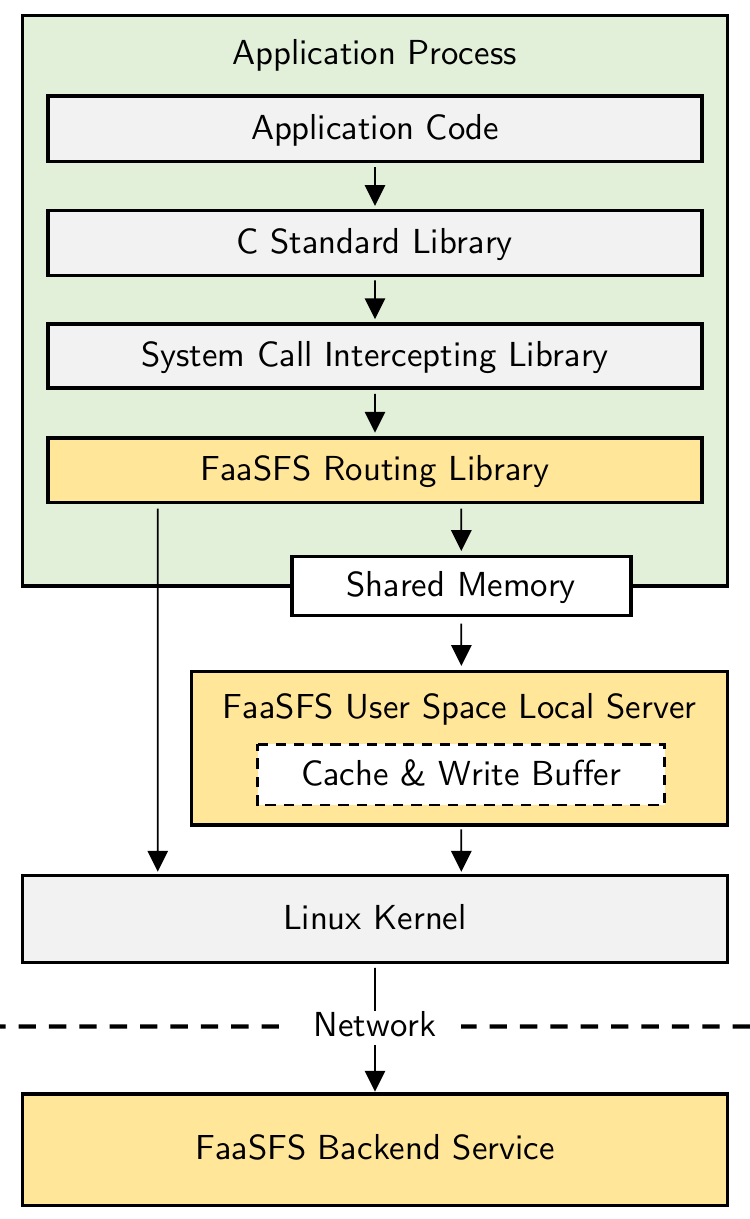}
  \caption{Overview of of \system.
  Yellow indicates the software that we wrote. }
  \label{fig:tsfs-components}
\end{figure}

Figure~\ref{fig:tsfs-components} shows the components of \system in the context of an application.
In the discussion that follows, we work our way down the stack from the application.
In our workloads, all system calls originate in a limited set of shared libraries: the C standard library, the pthread library, or the dynamic linker library.

\textbf{System call intercept:}
Operating in the AWS Lambda environment, we have no ability to modify or configure the kernel.
The system calls used by \texttt{ptrace} are also blocked, so we resort to binary modification via hot patching to intercept system calls.
We do this using the System Call Intercepting Library~\cite{syscallintercept} provided as part of the Persistent Memory Development Kit~\cite{Scargall2020}.

\textbf{Routing:}The \system Routing Library runs in the address space of the application and registers a handler with the System Call Intercepting Library.
This handler gets invoked ahead of all system calls, and can either let them pass unchanged or substitute an alternative implementation.
For those calls corresponding to the POSIX file system API it performs a routing decision, using arguments such as the path name or the file descriptor to determine whether the operation should go to \system or to the underlying operating system.
For paths, we test the prefixes, e.g., \texttt{/mnt/tsfs}, normalizing them to account for relative paths.
Some delicate bookkeeping is required, and in the case of forked processes this information must be carried in environment variables.

\textbf{Shared memory IPC:}
The Routing Library and the Local Server communicate using a shared memory area.
We maintain a set of buffers, configurable in number and size, to allow for concurrent requests.
By default we provide 10 buffers each 2~MB in size.
A client, which may correspond to either a thread or a process, first checks out a buffer using atomic operations.
It then writes the request data and marks it as ready for processing by the server.
The server will busy wait, spinning for up to 16~$\mu$s before falling back to wait on a semaphore.
The response works the same way, with the client first spinning in hopes of receiving a low-latency response, before turning to operating system support for coordination.
We analyze the efficiency of the IPC mechanism in Section~\ref{subsec:microbench}.

\textbf{User Space Local Server:} The \system User Space Local Server runs in a separate process from the application, and each instance of a cloud function runs one such process.
The Local Server maintains a cache as well as a write buffer, and intermediates all network communication with the \system Backend Service.
Both the Local serer and the Backend Service are written in Go, and we use Go's built-in RPC for communication between the two.
A further discussion of the transactional mechanisms follows in Section~\ref{subsec:consistencyimpl}.

\textbf{Backend Service} Our focus in this work is on the protocols necessary to maintain performance and POSIX-compliant consisstency across a large number of distributed cloud function.
For the purposes of this work we provide a prototype backend implemented as a monolithic server that maintains state in memory.
The techniques for building a scalable transactional backend service are well documented (e.g., ~\cite{cockroachdb,corbett2013spanner,verbitski2017amazon}) and we believe they can be combined with this work in the future.

\system comprises 3,500 lines of C and 15,000 lines of Go.
We delve deeper into the implementation of our caching and consistency protocols in the next section.

\subsection{Caching and Transactional Consistency}
\label{subsec:consistencyimpl}

\begin{figure}[!t]
    \centering
    \includegraphics[scale=1.0]{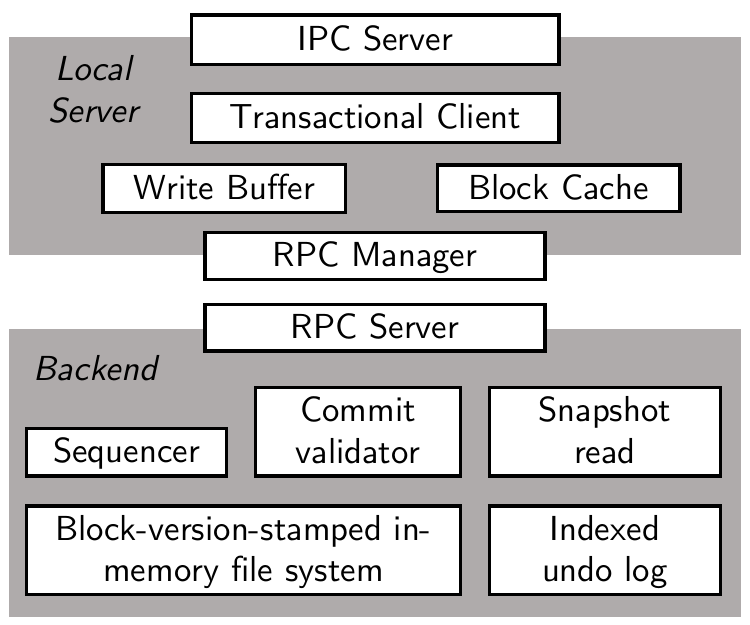}
    \caption{Components of the \system transactional implementation.}
    \label{fig:internalarchitecture}
  \end{figure}

We now turn to the transactional mechanisms implemented in \system. The diagram in Figure~\ref{fig:internalarchitecture} provides an overview of the major components comprising it.
As discussed in Section~\ref{subsec:posix}, we provide strict serializability with an optimistic concurrency control implementation.
At the beginning of each transaction, the client communicates with the server to obtain a filesystem-wide read timestamp $T_R$, which corresponds to the most recently committed version of the file system.
This introduces a round trip but allows us to guarantee strict serializability.
All reads must be returned to the application as if they had been issued at $T_R$, unless the transaction itself has modified the state being read, in which case the effect of its operations must be reflected.
Each file is represented as a number of blocks, each of which has an associated timestamp $T$ reflecting the commit time of it last change.

Throughout the course of a transaction, the Transactional Client at the Local Server maintains a read set $\mathbf{R}$ and a write set $\mathbf{W}$.
Each read occuring during the course of a transaction adds a record of the form (\textit{blocknum}, $T_R$) to $\mathbf{R}$.
Similarly, writes record of the form (\textit{blocknum}, \textit{changed data}) are added to $\mathbf{W}$, where \textit{changed data} is of the form (\textit{offset}, \textit{byte}[]) and can represent a partial update to the state of the block.
Associated with each read and write, we record an assertion on the length of the file, which is discussed in more detail below.

At commit time, the Transactional Client of the Local Server sends the $\mathbf{R}$ and $\mathbf{W}$ to the Backend Service, where they are validated according to the rules of optimistic concurrency control~\cite{bernstein1981concurrency}.
Each block $B$ stored in the Backend Service has an associated write timestamp $T^B_W$.
For each record in $\mathbf{R}$, the Backend Service verifies that $T^B_W$ $\le$ $T_R$.
If this verification fails for any block, then the transaction is aborted.
When verification succeeds, the Backend obtains a commit timestamp $T_W$ from the Sequencer.
For each write recorded in $\mathbf{W}$, it  copies the pre-commit state of the block to an Undo Log, then updates the block to apply the transaction.

As a consequence of POSIX semantics, each file read is also implicitly a read of the file length, as its result can depend on two things other writes to the range of bytes requested.
First, it depends on any truncate operations that specify a length less than the last byte requested.
It can also depend on writes that begin at an offset greater than the last byte requested, because POSIX zero-fills files that have gaps.
The file length can change frequently, but it typically grows more often than it shrinks.
In \system, operations in $\mathbf{R}$ also represent a predicate read on the file length, i.e., an assertion of $file length >= last byte read$.
Reads beginning beyond the end of the file return 0 bytes and add the assertion $file length <= first byte read$ to $\mathbf{R}$, whereas only those reads limited by reaching the end of file assert $file length = last byte read$.
The assertions, are also validated at commit time.

There is a large design space and extensive previous work on providing caching for distributed transactional clients~\cite{franklin1997transactional}.
One limitation of the cloud functions environment that \system must account for is that instances become frozen between invocations, and their cached state can only be updated once they are running again.
The Local Server typically contacts the Backend Service at the beginning of a transaction, requesting a record of any blocks that might have changed.
In the simplest implementation, the Backend Service checks the transaction log to see which blocks have changed since the Local Server cache was last updated, then sends all of them over.
This does not scale, and our protocol allows it to instead send either a block-level or a file-level invalidation, instead of broadcasting the updated data.
Additionally, \system allows the Backend Service to do nothing at all to update a cache, to leave it stale and rely on the commit mechanism to abort any transaction that reads it.
There is a large space of possible policies that might be implemented with this mechanism, a proper exploration of which is beyond the scope of this paper.
Our implementation includes a simple frequency-based heuristic that sends commonly fetched blocks and invalidates others.

Because of Local Server caches, the \system Backend Service does not need to do very much work to operate as a multiversioned database.
In many cases, the lag in updating caches means that multiversioning is possible with no server interaction at all.
When a client accesses a block that is not found in cache, it sends its read timestamp $T_R$ along with its request to fetch the block from the Backend Service, which subsequently uses the Undo Log to retrieve an older version of the block.

\section{Evaluation}
\label{sec:evaluation}

\subsection{Performance Characteristics}
\label{subsec:microbench}

\begin{figure*}[!t]
  \centering
  \includegraphics[scale=1.0]{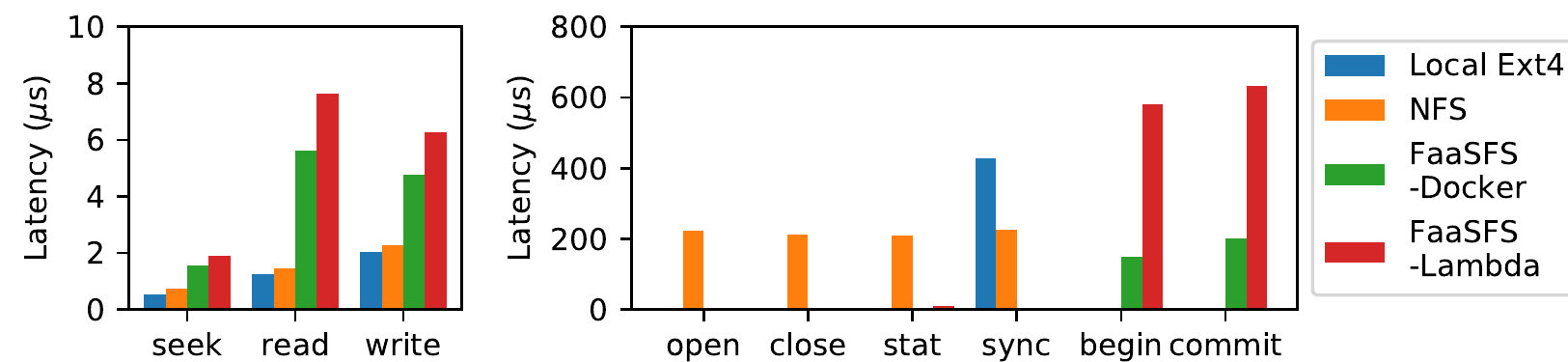}
  \caption{Median latency measurements show the overheads of our implementation of \system and the difference in remote access latencies.
  }
  \label{fig:latencies}
\end{figure*}

We begin our evaluation with a comparison of the per-operation latency of common operations in \system, which illustrates some consequences of our design.
The test loops over a sequence of operations on one file which it opens, reads at a random location, writes at a random location, syncs, and closes.
In the case of \system, these operations are wrapped in a transaction.

Figure~\ref{fig:latencies} shows the resulting latencies for a block size of 1~KB.
In order to facilitate comparisons to NFS, which is not available in AWS Lambda, we used a Docker image designed to replicate the AWS Lambda environment, running it on an EC2 instance (c5.large instance type).
Our back-end server is a c5.9xlarge instance, for both NFS and \system.
All of our environments use a Linux 4.14 kernel on both clients and servers.

The fastest operation is \texttt{seek}, which has a median latency of about 520~ns for a local file system, 750~ns for calls against an NFS target, and 1.6~$\mu$s for \system file handles on EC2.
Seek is a trivial operation, the performance of which  is dominated by system call overhead, or in our case, by the latency of our IPC implementation.
On Lambda this latency increases to 1.9~$\mu$s.

Our implementation of \system also pays a latency penalty on account of its IPC mechanism, which is slower than a native system call.
Ext4 and NFS pay a similar latency cost for reading 1~KB, requiring 1.25~$\mu$s and 1.4~$\mu$s, respectively, whereas our implementation of \system requires 5.6~$\mu$s on EC2 and 7.6~$\mu$s on Lambda.
When writing, Ext4 and NFS require 2.0~$\mu$s and 2.2~$\mu$s, whereas \system takes 4.8~$\mu$s on EC2 and 6.2~$\mu$s on Lambda.
Most of the added latency in \system comes from our IPC mechanism, which must copy the data two times.
\jmh{Justify that claim?} \jss{we can do a study to make sure we understand what is going on}
Reads take slightly longer than writes because our implementation of multiversion isolation incurs greater overheads on reads than on writes.

We believe that an implementation of \system as a kernel module could bring many of these overheads in line with the other implementations. \jmh{Would be easier to believe with justifications above.}
As discussed in Section~\ref{sec:design}, our user-space approach is driven by a desire to deploy \system in AWS Lambda and other cloud function platforms commercially available today.

\subsection{Filebench}


\begin{figure*}[!t]
  \centering
  \includegraphics[scale=1.0]{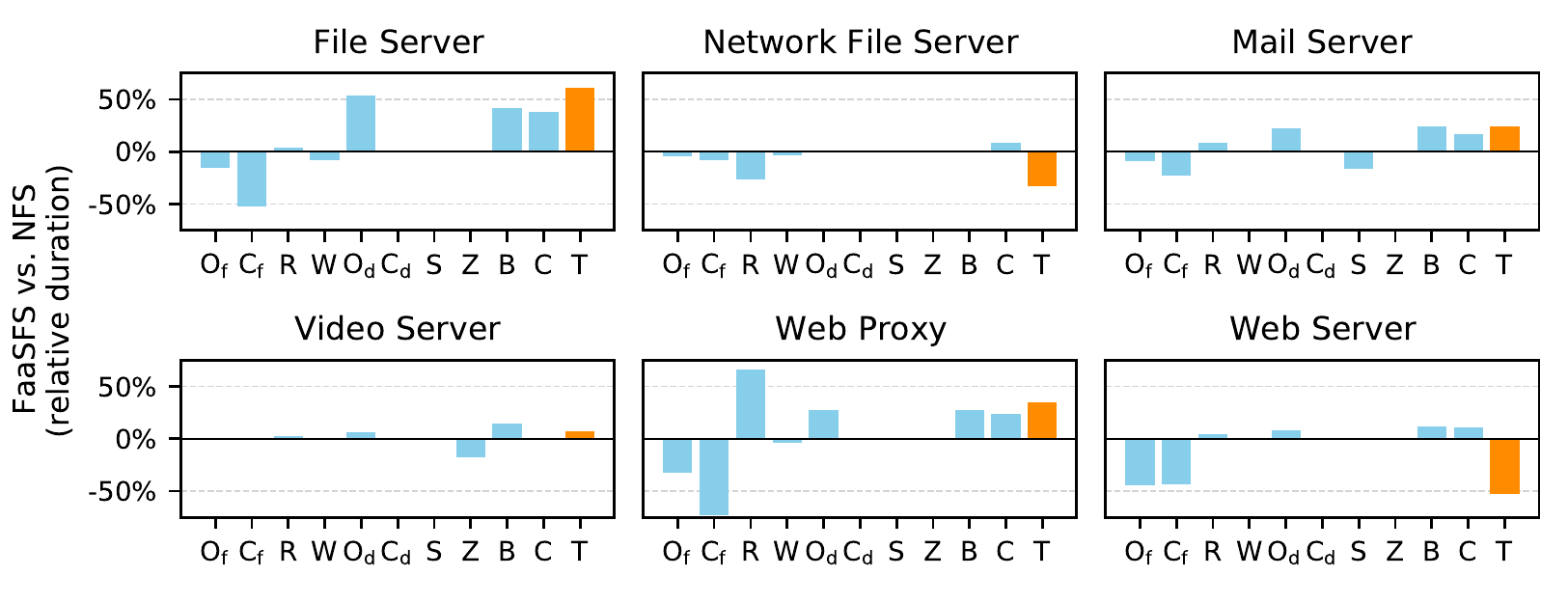}
  \caption{Filebench workload. Difference in average time elapsed between \system and NFS (lower is better).
  Columns are (O$_{\rm f}$) open file, (C$_{\rm f}$) close file, (R) read file, (W) write file, (O$_{\rm d}$) open directory, (C$_{\rm d}$) close directory, (S) fsync, (Z) rate limit, (B) begin, (C) commit, and (T) total overall.}
  \label{fig:applications}
\end{figure*}

To demonstrate the ability of \system to execute a variety of simulated applications, we used the Filebench~\cite{tarasov2016filebench} test suite.
We ran six of the standard included ``personalities'': file server, network file server, mail server, video server, web proxy, and web server.
These classic workloads represent a variety of i/o patterns, and thus provide a flavor for the diversity of applications that \system can support.
In adapting Filebench to the cloud functions setting, we wrap each iteration of the workload in a transaction.

Figure~\ref{fig:applications} compares \system running on EC2 to NFS with four concurrent Filebench clients.
For each operation, we plot the difference in the time spent during one iteration of the workload, then divide by the average iteration duration in the NFS base case ((\textit{\system{} op time} - \textit{NFS op time})/(\textit{NFS all ops time})).
This indicates how changes in the speed of each operation impact overall benchmark performance.
In the last column we also show this overall performance difference, which is the cumulative total of the differences for each operation.
\jmh{These charts are really nice breakdowns but the changing denominator per bar means we can't understand why the orange bars look the way they do. It would also be really nice to see P99s or some other representation of variance.} \jss{the denominator is the same for each bar in the box, different between boxes. I've tried to clarify.}

Our implementation of \system outperforms NFS in some workloads, but lags it in others.
For example in the file server workload, \system gains significant advantages from faster file open and close operations (O$_{\rm f}$ and C$_{\rm f}$), but pays a penalty when opening directories (O$_{\rm d}$), beginning transactions (B), and committing (C).
Overall it is 61\% slower.
The web server workload, by contrast, has wins and losses on the same operations, but overall runs about 2.1x faster.
This discrepancy is driven primarily by the number of operations executed in each transaction, which is 3x greater for the web server than it is for the file server.
The network file server gains a large advantage in read operations (R), and an overall win in performance, attributable to more effective caching in \system.
With the web proxy, by contrast, \system sees a disadvantage for read operations (R), this time attributable to the increased overhead of accessing cached data.
For the mail server, we note the reduction in time spent in sync operations (S), though this does not outweigh the cost of time spent in begin (B) and commit (C).
In the video server, Filebench implements a per-client rate limit.
Here we see that the much of the time added in begin and commit gets absorbed the rate limit (Z), so the overall performance impact is minimal.

\subsection{TPC-C with SQLite}
\label{subsec:tpcc}

\begin{figure}[!t]
  \centering
  \includegraphics[width=3.3in]{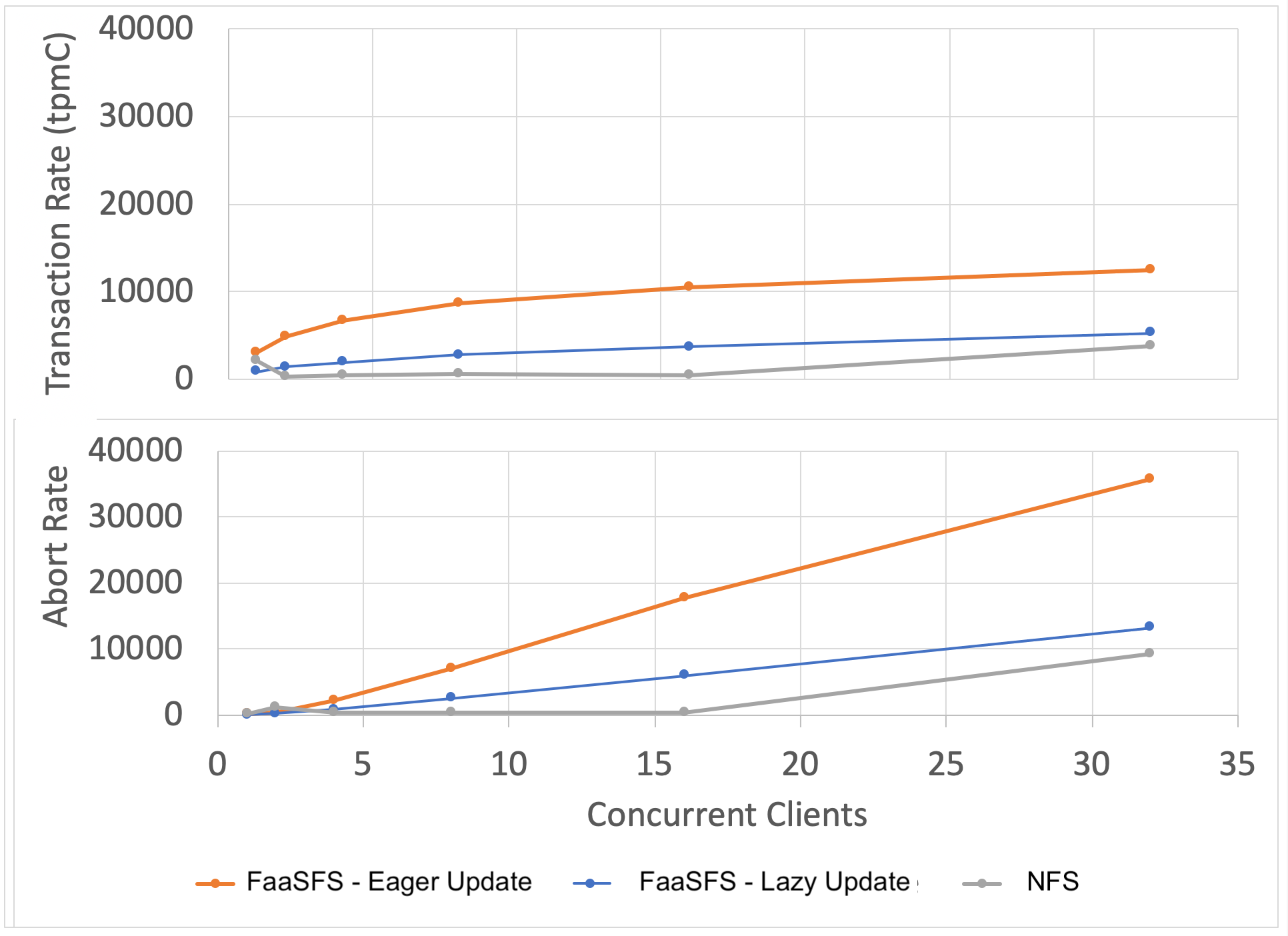}
  \caption{TPC-C scaling. We compare \system to NFS and consider two configurations, one in which caches are updated eagerly for all files, and the other in which the cache for each file is updated the first time the transaction accesses it.}
  \label{fig:tpcc}
\end{figure}

When considering workloads that challenge \system, it is hard to come up with something more demanding than an OLTP database.
We chose the TPC-C benchmark~\cite{tpcc} to explore the limits of what \system could achieve under a contended, write-heavy workload with strict correctness requirements.
Whereas Filebench issues operations without confirming that they do the right thing, a database will quickly detect corruption if the underlying file system does not live up to the POSIX guarantees.

We chose to run TPC-C on SQLite~\cite{sqlite}, a database that runs as a library and persists its state in a file system.
SQLite is primarily designed for embedded environments, and aims to meet high standards for efficiency and management-free operation.
The target environments may have limited operating system facilities, e.g., they may not support shared memory, and by default SQLite communicates through the file system when coordinating among multiple processes.
SQLite does only limited caching itself, relying heavily on caching in the underlying file system to achieve good performance.
Its page cache is maintained on a per-session basis, and gets cleared any time the database file changes.
SQLite allows only one writer to access the database at a time, however it supports multiple readers.
Readers may run concurrently with one another, and concurrently with writers when multiversion concurrency control is enabled.
In most cases one can think of a SQLite database as a single file, however SQLite allows an application to open multiple such database files, and can perform transactions spanning them using a two phase commit protocol~\cite{gray1992transaction}.

The TPC-C workload models the operation of a large company with many regional warehouses and geographically distributed customers.
The database serves queries for customers inspecting stock and placing orders, as well as for processing payments, and tracking deliveries.
It is a write-heavy workload, with about 70\% of queries resulting in a modification to the database.
There is some locality of reference, since 90\% of orders are served entirely from the customer's regional warehouse, whereas 10\% include one or more items from another warehouse.
This means when partitioned by warehouse, many queries can complete locally.

In our configuration we use 64 warehouses, split across 64 SQLite database files, and a total database size of 756~MB.
Figure~\ref{fig:tpcc} compares performance on NFS to two configurations of \system.
In the \textit{eager} configuration client caches receive updates for all changes to the file system at the beginning of the transaction, whereas in the \textit{lazy} configuration \system updates cached data for each file only when it is opened.

This experiment highlights both strengths and weaknesses of our approach: there is a significant improvement over NFS, nearly 30x in some cases, however the fraction of transactions aborted due to concurrent modification rises rapidly.
When there is only one client active NFS and eager \system  have comparable performance, with 2,500 and 3,000~tpmC, respectively.
With two clients NFS performance is degraded by a factor of 10, as clients must invalidate an entire cached file whenever any part of it changes.
\system, by contrast, ships sends changed blocks rather than invalidating caches, and sees a 70\% increase in performance when going from one client to two.
The eager update policy in \system sends changes for all files at the beginning of each transaction, whereas the lazy update policy defers fetching changes until the file is accessed.
While the eager policy improves throughout 2-3x it uses significant amounts of network bandwidth and server resources.
To ensure that back end capacity would not be a bottleneck, we used a \texttt{c5.18xlarge} EC2 instance.

While this experiment demonstrates the completeness of our \system implementation and highlights some of its scaling characteristics, we do not advocate running a write-intensive application like that modeled by TPC-C using the combination of SQLite and \system.
One potential objection is that layering one transactional system on top of another is bound to be inefficient, however SQLite helpfully provides a mode that turns off all crash recovery mechanisms.
A more serious problem is that SQLite maintains a sequence id that is updated with every transaction, so even though \system can elide locks it is unable to support concurrent updates to a SQLite database file.
This represents false sharing, where transactions that otherwise operate on disjoint sets of database blocks nonetheless contend for the one that records the database version.
We come back to how this might be addressed in Section~\ref{sec:futurework}.

A back-of-the-envelope calculation suggests that the cost of running TPC-C using Lambda and \system is comparable to that of server based implementations.
According to published commercial benchmarks~\cite{tpccresults}, the capital cost of a database is approximately \$1 USD per tpmC, which works out to $\$0.60$ per million transactions when amortized over 3~years.
By way of comparison, a cloud function in AWS Lambda with 1~GB memory costs \$0.001 per minute, which based on our experiments suggests a cost of $\$1.00$ per million transactions.
Neither of these represent an all-in cost.
Cost figures for commercial systems leave out power and other data center costs, and assume 100\% utilization, whereas our \system estimate includes the cost of Lambda but not the cost of providing the back end and storage.
For suitable workloads, those that are read-heavy and have limited write contention, running an existing database like SQLite on top of \system could be a  practical solution.
Interestingly, the SQLite authors described it as ``serverless database'' before serverless was used in the cloud context~\cite{sqliteserverless}.
Perhaps \system can turn the world's most widely deployed embedded database into a viable cloud database.


\subsection{Full-stack application}
\label{subsec:fullstack}


\begin{figure}[!t]
  \centering
  \includegraphics[scale=1.0]{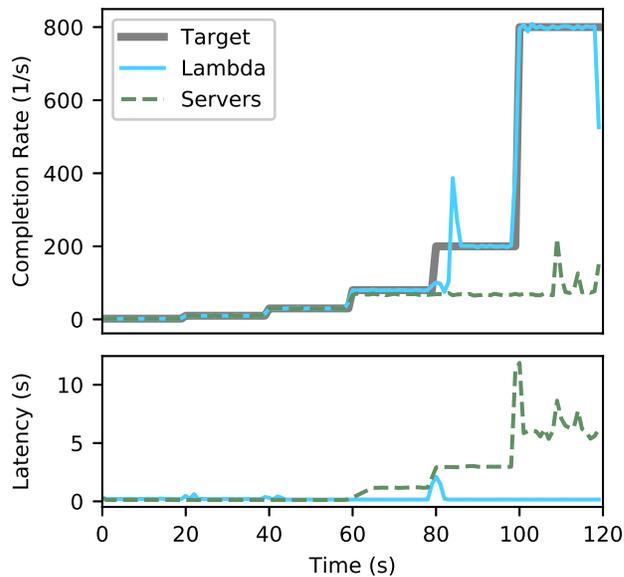}
  \caption{A blog application backed by \system on AWS Lambda achieves serverless scaling, whereas a 2-server configuration which may have ample capacity most of the time does not scale to handle a load spike. In this experiment, the number of clients increases over time, with a target given by the gray line.}
  \label{fig:lambdagscaling}
\end{figure}


\begin{table}[tbp]
\centering
\begin{tabular}{|l|l|r|}
\hline
                            & \multicolumn{1}{|c|} {Component}                 & \multicolumn{1}{|c|} {Cost per million} \\
\hline \hline
\multirow{3}{*}{Serverless} & API Gateway               & \$3.50                      \\
                            & Lambda                    & \$5.21                    \\
\cline{2-3}
                            & Total                     & \$8.71 \\
\hline \hline
\multirow{3}{*}{Servers}    & Load Balancer & \$0.12                    \\
                            & 2x EC2 m5.large           & \$0.76 \\
\cline{2-3}
                            & Total                     & \$0.88 \\
\hline
\end{tabular}
\caption{Pricing breakdown and comparison for blog application. The per-request cost of serverless could be up to 10x that of fully-loaded servers.
As we dicsuss in section~\ref{subsec:fullstack}, several factors are likely to reduce the difference in practice.}
\label{fig:pricing}
\end{table}

In order to understand how \system performs in a full-stack application we chose to evaluate it for running Mezzanine~\cite{mezzanine}, a popular open source blogging platform written in Python.
Mezzanine is a web application adhering to Python's WSGI standard, and we were able to deploy it to AWS Lambda using Zappa~\cite{zappa}, an open source project that packages traditional WSGI applications as cloud functions.
Zappa uses AWS API Gateway to make cloud functions accessible over the internet.
The resulting application, including libraries, is approximately 100~MB uncompressed, 30~MB compressed.
In addition to the application payload, the cloud function includes a custom runtime, provided as a Lambda layer, that is 134~MB uncompressed, 62~MB compressed.
We run Mezzanine in its default storage configuration, which uses a SQLite~\cite{sqlite} database.

For comparison, we provisioned a cluster of two \texttt{m5.large} EC2 instances, placing them behind an AWS Application Load Balancer for load balancing, and connecting them to a shared AWS Aurora MySQL database.

Figure~\ref{fig:lambdagscaling} shows a workload where a number of clients repeatedly load the home page of the blog with a target rate of one request per second.
To simulate remote clients we run the load generator on AWS infrastructure in a different region (approximately 90~ms round-trip network latency).
We ramp up the number of clients from 2 to 800, measuring service latency and throughput.
While the two servers are able to sustain 70 requests per second, AWS Lambda with \system adds resources to meet much greater demand, showing only brief latency spikes as it provisions more resources.
We were not able to identify an impact to latency from making simple updates (adding blog entries, posting comments).

At low load, average latency with servers is 122~ms whereas with Lambda it is 207~ms.
We believe that this is mostly caused by the overheads of API Gateway and Lambda invocation.
Note that as the service warms up it gets faster: running 1,200 concurrent clients we average 7,200 requests/sec with a latency of 156~ms.

This test is designed to illustrate one compelling use case for \system, making it possible to host a small web site on a pay-per-use basis, so usually very little cost, while maintaining the ability to reach large scale quickly when needed.

Figure~\ref{fig:pricing} shows a breakdown of the costs for the serverless and server-based implementations.
This makes it appear that serverless is 10x more expensive than servers, but in practice this is unlikely to be the case.
First, one must maintain and pay for idle capacity on the servers, both in case of failure and in case of load spikes.
For a two-server configuration, fault tolerance requires a 2x over-provisioning, and intra-day peak-to-average skew can add another 2x.
The cost of serverless can also be reduced by replacing API Gateway with an Application Load Balancer, as used with EC2 (though this is presently not supported by Zappa).
We may be able to further reduce the cost of using \system on Lambda by optimizing our IPC mechanism, which spins the CPU aggressively.
An alternative implementation might allow us to achieve similar performance using cheaper functions configured with less memory (high-memory includes high CPU in Lambda).
\jmh{it's a particular weird analysis as you didn't provision the servers for your peak workload, and you don't measure any kind of penalty for underperforming. 
If you provisioned the servers for peak, you'd pay something like 10x more for them to get to 700 requests/sec. That's still no bargain for serverless, given that it's mostly idle, but it's probably more realistic -- it's what you'd do if underperforming was really expensive in terms of real-world costs..}
\jss{we'll come back to this}


\section{Related Work}
\label{sec:relatedwork}

\jss{Need to add Rethink the Sync}
\jss{Should add CAPFS}

\subsection{State in serverless computing}

A number of research efforts have recognized that serverless computing is a unique environment with state management needs that remain unmet.
The Anna key-value store focuses on the elastic scalability demands of a serverless environment, demonstrating performance over a wide range of scale, and automatic storage tiering that adapts to application needs~\cite{wu2018eliminating,wu2019anna}.
Anna is extended by Cloudburst~\cite{Sreekanti2020-ec}, which adds a FaaS execution layer with integrated caching of key-value store data.
Pocket~\cite{klimovic2018pocket} provides ephemeral storage for serverless analytics, focusing on efficient resource allocation for short time durations, a challenge also studied by Locus~\cite{pu2019shuffling}.
AFT addresses provides an atomicity shim that sits between cloud functions and cloud storage~\cite{sreekanti2020fault}, which makes it easier to achieve the idempotence that cloud functions require.
Serverless computing owes its popularity in part to compatibility with existing software ecosystems, and \system is unique in adapting POSIX APIs to cloud functions.

\subsection{Shared file systems}

Shared file systems originated with NFS~\cite{sandberg1985design} and have subsequently been subject to extensive research.
A consistent theme in this work has been achieving both consistency and performance, with notable work including Coda~\cite{howard1988scale}, Sprite~\cite{nelson1988caching}, and V~\cite{gray1989leases}.
Ideas from this work have found their way into into contemporary protocols, including NFSv4~\cite{nfs4rfc} and SMB~\cite{smbprotocol}, however these systems are still subject to the limitations discussed in Section~\ref{subsec:sharedfilesystems}, as they fundamentally rely on locks, leases, and write-through caching to provide consistency with shared state.
Variations on this theme occur in the high-performance computing space, where Lustre~\cite{schwan2003lustre} is popular, and incorporates intent based locks that allow write back caching.
Another category of shared file systems includes cluster file systems like GFS~\cite{Preslan1999-xx} and OCFS2~\cite{Fasheh2006-nr}.
These file systems assume that all participants have access shared block storage, but this makes them vulnerable to misbehaving clients, and they are thus not candidates for cloud storage.
In the cloud context shared file system research has focused on back end scalability, especially for metadata, e.g., in Ceph~\cite{weil2006ceph}, a well as extreme scale, as in Google File System~\cite{ghemawat2003google,dean2010evolution}, HDFS~\cite{shvachko2010hadoop}.
Delta Lake~\cite{deltalake} is a recent transactional shared file system designed specifically for analytics workloads.
However, it does not offer POSIX semantics.

\subsection{Transactional file systems}

QuickSilver~\cite{schmuck1991experience,haskin1988recovery} is perhaps the most direct predecessor to \system.
It provides operating system support for distributed transactions.
Similar to transactions transparently delineated by cloud function boundaries in \system, QuickSilver creates a per-process \textit{default transaction} if none was specified explicitly.
However, QuickSilver implements blocking transactions rather than using optimistic concurrency control, provides weaker isolation, and does not implement client-side caching.
It does not explore any of the performance boosting elements of this work.

The Inversion file system~\cite{olson1993design} built on top of POSTGRES~\cite{stonebraker1986design} maps directory data and file blocks to relational tables, inheriting the isolation guarantees of the underlying database.
Its POSIX compatibility is limited as it only implemented a small number of basic operations, while adding transactional extensions.
Inversion benefits from caching at the backend database server (via the buffer pool), but it doesn't incorporate any client-side caching.

Various other efforts have sought to combine file systems and databases.
Informix patented the idea of providing a file system API atop a database backend~\cite{balabine1999file}, and since 2006 Microsoft has shipped TxF a non-shared and now deprecated~\cite{mstxf} transactional file system with Windows.
Transactional file system APIs designed to make crash recovery simpler have been provided in AvdFS~\cite{Verma2015-ij}, CFS~\cite{Min2015-ql}, TxFS~\cite{Hu2019-lx}, and TxOS~\cite{Porter2009-vq}.
However these systems do not consider the distributed setting.

\subsection{Local caching in transaction systems}

A key challenge we encountered in \system is how state makes its way to client caches.
The database community has studied this problem, especially in the context of object databases, and a body of work has been surveyed, compared, and categorized~\cite{franklin1997transactional}.
There are also middle-tier caching accelerator implementations such as Ganymed~\cite{plattner2004ganymed}, which routes read transactions to replicas, and MTCache~\cite{larson2004mtcache}, an extension to SQL Server, which provides semantics equivalent to executing transactions at the database server, but with the twin benefits of offloading work from the centralized system and lowering latency.
MTCache relies on materialized views, similar to the approach used by TimesTen~\cite{lahiri2013oracle}.
This work might provide useful approaches to updating client caches in \system.
One notable recent system is Sundial~\cite{yu2018sundial}, which is particularly close in spirit to our work since it uses optimistic concurrency control and integrates this concurrency control mechanism with its caching mechanism, as we do.
Sundial promises improved concurrency, but more work is needed to determine whether its approach can be reconciled with the consistency needs of POSIX workloads.

\section{Future Work}
\label{sec:futurework}

\system as implemented today is a prototype with a monolithic in-memory back-end.
This has been appropriate for validating our design choices and testing them in the the cloud functions environment, including caching, concurrent transaction conflict resolution, yet it remains a research artifact.
Future work can rely on proven techniques for building scalable and distributed transactional systems~\cite{corbett2013spanner,stonebraker2013voltdb,cockroachdb} to create a system that can deployed in practice.

This work has focused on the mechanisms of optimistic transactions, but leaves open a number of policy questions.
Cache update policy is one area in which there have been many proposals~\cite{franklin1997transactional}.
We are also interested in exploring more optimal techniques for providing strict serializability e.g., with loosely synchronized clocks~\cite{adya1995efficient}.

The choice of consistency model is also open to exploration.
While we chose to implement serializability with external consistency, but some applications may run correctly with weaker guarantees.
The optimistic implementation of snapshot isolation~\cite{berenson1995critique} is similar to that for serializability, and in some cases this might suffice.
An interesting question in this context is under what circumstances lower isolation still allows lock elision, which is import for some applications, but which others might choose to forego.
Weak consistency models, including eventual consistency methods~\cite{bailis2013eventual,shapiro2011conflict,decandia2007dynamo,conway2012logic} offer an alternative approach~\cite{brewer2012cap}, and may offer a more appropriate set of guarantees for some applications.

\section{Conclusion}
\label{sec:conclusion}

We report on our experience developing \system, a shared file system designed specifically to meet the needs of serverless applications.
\system brings together the widely-used POSIX filesystem API with the
operational and economic benefits of serverless computing. 
This demonstrates the potential to
broaden the scope of serverless computing to encompass some of the most popular
applications in use today, and to welcome developers who are used to a traditional file system API.



The design of \system exploits the fact that cloud functions are finite in duration, 
with clear begin and end points. This allows us to transparently use transaction mechanisms to achieve 
consistency guarantees that POSIX applications expect. Transactions also match the cloud function model of reliability, in 
which function executions are the unit of fault tolerance. Atomic commit, coupled with automatic retry of 
idempotent functions, gives exactly-once semantics.

While transactional file systems traditionally pay a performance penalty,
we show that optimistic concurrency control can make transactions perform well in the serverless file system context.
Our evaluation shows that the performance of \system is usually comparable to that of traditional file systems, and that our POSIX implementation is sufficiently complete and compliant to run a variety of application benchmarks, as well as a full-stack web application.
We hope that \system will help open up serverless computing to a broader range of applications.

\bibliographystyle{abbrv}
\bibliography{references}

\clearpage

\end{document}